\documentclass[preprintnumbers,amsmath,amssymb]{revtex4}
\usepackage{amsmath, amssymb}

\newcommand{\commentoutA}[1]{}

\usepackage{graphicx}
\begin{document}

\preprint{LA-UR-14-26535}

\title{Generalized extended Lagrangian Born-Oppenheimer molecular dynamics}

\author{Anders M. N. Niklasson} 
\email{amn@lanl.gov}
\author{Marc J. Cawkwell}
\affiliation{Theoretical Division, Los Alamos National Laboratory, Los Alamos, New Mexico 87545}

\date{\today}

\begin{abstract}
Extended Lagrangian Born-Oppenheimer molecular dynamics based
on Kohn-Sham density functional theory is generalized in the limit of vanishing 
self-consistent field optimization prior to the force evaluations.
The equations of motion are derived directly from the extended Lagrangian
under the condition of an adiabatic separation between 
the nuclear and the electronic degrees of freedom. We show how this
separation is automatically fulfilled and system independent.
The generalized equations of motion require only one diagonalization 
per time step and are applicable to a broader range of materials 
with improved accuracy and stability compared to previous formulations. 
\end{abstract}

\maketitle

\section{Introduction}
Born-Oppenheimer molecular dynamics simulations,
where classical molecular trajectories are propagated by forces 
that are calculated on-the-fly from the relaxed electronic ground state in each time step,
provide a powerful tool in materials science, chemistry and biology \cite{DMarx00,BKirchner12}. 
However, the computational cost is high.
In regular Born-Oppenheimer molecular dynamics simulations based on Kohn-Sham
density functional theory \cite{hohen,KohnSham65,RParr89,RMDreizler90}, 
the cost scales with the cube of the number of atoms, 
which is multiplied by a significant prefactor given by the number of
self-consistent field iterations that are required to find the 
relaxed electronic ground state prior to the force evaluation in each time step. 
If a sufficient degree of convergence is not achieved, 
the electronic system behaves like a heat sink or
source, gradually draining or adding energy to the atomic system due to a broken
time-reversibility in the propagation of the underlying electronic degrees
of freedom \cite{DRemler90,PPulay04,JMHerbert05,ANiklasson06,TDKuhne07}.
Recently, an extended Lagrangian formulation of Born-Oppenheimer molecular
dynamics was proposed that overcomes this problem 
\cite{ANiklasson08,PSteneteg10,GZheng11,JHutter12,LLin14,PSouvatzis14}.
By including an auxiliary electron density as a dynamical variable, the
equations of motion can be integrated without breaking time reversibility, 
even under approximate self-consistent field convergence. This framework 
enables Born-Oppenheimer molecular dynamics simulations \cite{MCawkwell12,DBowler14}
with accurate long-term energy conservation using fast linear scaling solvers 
\cite{SGoedecker99,ANiklasson02,DBowler12} 
that can treat much larger systems than previously possible.

The extended Lagrangian formulation 
was recently investigated in the limit of vanishing self-consistent field convergence
prior to the force calculations. This fast approach requires only one 
diagonalization per time step \cite{ANiklasson12,PSouvatzis13,PSouvatzis14}. 
In this limit, the iterative ground state optimization, which is 
necessary in regular Born-Oppenheimer molecular dynamics, 
is replaced by a {\em dynamical} optimization acting over time \cite{PBendt83,RCar85}. 
However, this fast self-optimizing dynamics relies on the ability to keep the electronic
degrees of freedom close to the exact self-consistent ground state, which is
approximated using a simple linear mixing.
For many materials systems exhibiting, for example, bond breaking or charge sloshing, this
linear mixing is insufficient and leads to instabilities. In this article we revisit extended Lagrangian
Born-Oppenheimer molecular dynamics in the limit of vanishing self-consistent field optimization
and propose a more general formulation that is applicable to a broader range of materials.
In contrast to previous formulations we also derive the equations of motion directly from the
extended Lagrangian without intermediate steps. 
This is possible in an adiabatic limit under the condition of a frequency separation
between the nuclear and the electronic degrees of freedom, which we show is automatically fulfilled.

We present our generalized extended Lagrangian Born-Oppenheimer molecular dynamics in terms of
Kohn-Sham density functional theory \cite{hohen,KohnSham65,RParr89,RMDreizler90}
and we use the electron density as the variable
for the electronic degrees of freedom. However, the framework should 
be generally applicable also to other electronic structure formulations, 
for example, density matrices in Hartree-Fock calculations and
wavefunctions or Green's functions in methods for strongly correlated electrons. 

\section{Extended Lagrangian Born-Oppenheimer molecular dynamics}

\subsection{Generalized extended Born-Oppenheimer Lagrangian}

Our generalized extended Born-Oppenheimer Lagrangian is defined by
\begin{equation}\label{XLBO}\begin{array}{l}
{\displaystyle {\cal L}({\bf R},{\bf \dot R}, n, {\dot n}) =   
\frac{1}{2}\sum_I M_I{\dot R}^2_I - {\cal U}\left({\bf R},n\right)
+ \frac{1}{2}\mu \int{\dot n}^2({\bf r})d{\bf r}}\\
~~\\
{\displaystyle -\frac{1}{2}\mu {\omega}^2\iiint\left(\rho({\bf r}_1)-n({\bf r}_1)\right)^\dagger
K^\dagger({\bf r}_1,{\bf r}_2)K({\bf r}_2,{\bf r}_3)
\left(\rho({\bf r}_3)-n({\bf r}_3)\right) d{\bf r}_1 d{\bf r}_2 d{\bf r}_3 }.
\end{array}
\end{equation}
The first term is the kinetic energy for the nuclear degrees of freedom, 
i.e.\ the atomic coordinates ${\bf R} = \{R_I\}$ and the velocities ${\bf {\dot R}} = \{{\dot R}_I\}$, where
the dot denotes the time derivative.
The second term is a modified Kohn-Sham Born-Oppenheimer potential energy, ${\cal U}\left({\bf R},n\right)$, 
including nuclear-nuclear repulsion. A more precise definition of ${\cal U}\left({\bf R},n\right)$
is given in Sec.\ \ref{Sec2B} below.
The last two terms form a harmonic oscillator for the extended electronic degrees of freedom, i.e.\ the
electron density $n({\bf r})$ and its time derivative ${\dot n}({\bf r})$. Here $\mu$ is an electron
mass parameter and $\omega$ is the frequency determining the curvature of the harmonic well
centered around the optimized density $\rho({\bf r})$. The electron density $\rho({\bf r})$
is a fully relaxed and optimized ground state density, but for an approximate 
linearized electron-electron interaction. Details about the definition of 
$\rho({\bf r})$ are given below in Sec.\ \ref{Sec2B}.
In contrast to the previous formulation of extended Lagrangian
Born-Oppenheimer molecular dynamics \cite{ANiklasson08}, the harmonic potential includes 
a non-local kernel $K({\bf r,r'})$. This kernel is defined by the relation
\begin{equation}\label{Kernel}
{\displaystyle \int K({\bf r,r'}) \left(\frac{\delta \rho({\bf r'})}{\delta n({\bf r''})} 
- \delta({\bf r'-r''})\right) d{\bf r'} = \delta({\bf r-r''}) }.
\end{equation}
A more detailed explanation for this particular choice, which forces the dynamical variable $n({\bf r})$
to evolve close to the exact fully optimized ground state density, 
is given in Sec.\ \ref{SecKernel} below. However, a constant density-independent kernel 
can also be used. 

\subsection{Modified Born-Oppenheimer potential energy surface}\label{Sec2B}

To construct the modified potential energy surface ${\cal U}\left({\bf R},n\right)$ in Eq.\ (\ref{XLBO}), we start with
the Kohn-Sham Born-Oppenheimer potential energy surface $U({\bf R})$
\cite{KohnSham65,RParr89,RMDreizler90}, which is given by a constrained functional minimization,
\begin{equation}\label{DFTPot}
{\displaystyle U({\bf R}) =
\min_{\widetilde \rho} \left\{  E({\bf R},{\widetilde \rho})
~ \left\lvert ~ {\widetilde \rho}({\bf r}) 
= \sum_{i \in {\rm occ}} \left|\psi_i({\bf r})\right|^2, ~ \langle\psi_i|\psi_j\rangle = \delta_{ij}\right\}\right. + V_{\rm nn}({\bf R}) },
\end{equation}
where
\begin{equation}\label{DFTPotE}
{\displaystyle E({\bf R},{\widetilde \rho}) =
 \sum_{i \in {\rm occ}} -\frac{1}{2} \langle\psi_i|\nabla^2|\psi_i\rangle
+ \int V_{\rm ext}({\bf R}, {\bf r}){\widetilde \rho}({\bf r}) d{\bf r}
+ E_{\rm ee}[{\widetilde \rho}].}
\end{equation}
The first term in Eq.\ (\ref{DFTPotE}) is the kinetic Kohn-Sham energy for an electron density ${\widetilde \rho}({\bf r})$ that is given from
an occupied (occ) set of orthonormalized orbitals $\{\psi_i\}$.
The second term contains the external or pseudopotential potential, $V_{\rm ext}({\bf R},{\bf r})$, 
i.e.\ the electron-ion interaction, and the third term is the electron-electron interaction energy, 
\begin{equation}\label{DFTEE}\begin{array}{l}
{\displaystyle E_{\rm ee}[{\widetilde \rho}] =
\frac{1}{2}\iint\frac{{\widetilde \rho}({\bf r}){\widetilde \rho}({\bf r'})}{|{\bf r}-{\bf r'}|}d{\bf r}d{\bf r'} 
+ E_{\rm xc}[{\widetilde \rho}] },\\
\end{array}
\end{equation}
consisting of the Hartree term and the exchange-correlation energy. 
The $V_{\rm nn}({\bf R})$ term in Eq.\ (\ref{DFTPot}) is the ion-ion repulsion energy.
The constrained minimization in Eq.\ (\ref{DFTPot}) is performed over all normalized electron densities, 
representable by orthonormal orbitals. 
The minimization defines a ${\widetilde \rho}$-independent potential energy surface $U({\bf R})$, which
can be calculated through a non-linear iterative
self-consistent field optimization procedure that involves solving a sequence of
Kohn-Sham eigenvalue equations. The exact self-consistent (sc) ground state 
density, $\rho_{\rm sc}({\bf r})$, is the density for which the minimum $U({\bf R})$ 
is obtained in Eq.\ (\ref{DFTPot}).

The modified Born-Oppenheimer potential energy surface, ${\cal U}({\bf R},n)$ in Eq.\ (\ref{XLBO}), 
is constructed as $U({\bf R})$ if we approximate the electron-electron interaction term $E_{\rm ee}[{\widetilde \rho}]$
with a functional linearization around the electron density $n({\bf r})$, 
\begin{equation}\label{LinEE}\begin{array}{l}
{\displaystyle E^{(1)}_{\rm ee}[{\widetilde \rho},n]  = E_{\rm ee}[n] +
\left. \int\frac{\delta E_{\rm ee}[{\widetilde \rho}]}{\delta {\widetilde \rho}}\right\rvert_{{\widetilde \rho}=n}
\left({\widetilde \rho}({\bf r})-n({\bf r})\right)d{\bf r}}.
\end{array}
\end{equation}
Our modified Born-Oppenheimer potential energy surface is then given by the 
constrained minimization,
\begin{equation}\label{ShadowPot}
{\displaystyle {\cal U}({\bf R},n) =
\min_{\widetilde \rho} \left\{  E^{(1)}({\bf R},{\widetilde \rho},n)
~ \left\lvert ~ {\widetilde \rho}({\bf r}) 
= \sum_{i \in {\rm occ}} \left|\psi_i({\bf r})\right|^2, ~ \langle\psi_i|\psi_j\rangle = \delta_{ij}\right\}\right. + V_{\rm nn}({\bf R}) },
\end{equation}
where the electron-electron interaction in Eq.\ (\ref{DFTEE}) has been replaced by the linearized
expression in Eq.\ (\ref{LinEE}), i.e.
\begin{equation}\label{DFTPotLinE}
{\displaystyle E^{(1)}({\bf R},{\widetilde \rho},n) =
 \sum_{i \in {\rm occ}} -\frac{1}{2} \langle\psi_i|\nabla^2|\psi_i\rangle
+ \int V_{\rm ext}({\bf R}, {\bf r}){\widetilde \rho}({\bf r}) d{\bf r}
+ E^{(1)}_{\rm ee}[{\widetilde \rho},n].}
\end{equation}
The density $n({\bf r})$ appears in ${\cal U}({\bf R},n)$ as a dynamical variable in the same way as ${\bf R}$.
A given set of dynamical variables, $n({\bf r})$ and ${\bf R}$, determine the relaxed optimized ground state density
$\rho({\bf r})$, which is attained at the minimum that defines ${\cal U}({\bf R},n)$ in Eq.\ (\ref{ShadowPot}),
in the same way as the non-linear iterative optimization of regular Kohn-Sham density functional theory determines
the exact self-consistent ground state density $\rho_{\rm sc}({\bf r})$ in Eq.\ (\ref{DFTPot}), i.e.
\begin{equation}\label{Shadow_RhoMin}
{\displaystyle \rho({\bf r}) = \arg \min_{\widetilde \rho} \left\{  E^{(1)}({\bf R},{\widetilde \rho},n)
~ \left\lvert ~ {\widetilde \rho}({\bf r})
= \sum_{i \in {\rm occ}} \left|\psi_i({\bf r})\right|^2, ~ \langle\psi_i|\psi_j\rangle = \delta_{ij}\right\}\right. }.
\end{equation}
The main difference is that
the constrained minimum of the linearized form in Eq.\ (\ref{ShadowPot}) or (\ref{Shadow_RhoMin}) can be obtained from a single solution 
of a regular Kohn-Sham eigenvalue equation with the Kohn-Sham Hamiltonian $H_{\rm KS}[n]$ calculated at density $n({\bf r})$, i.e.
\begin{equation}\label{OneStep}
\rho[n]({\bf r}) = \sum_{i \in \rm occ} |\psi_i({\bf r})|^2, ~~~{\rm where} ~~~ H_{\rm KS}[n]\psi_i({\bf r}) = \varepsilon_i\psi_i({\bf r}).
{\displaystyle }
\end{equation}
To highlight this direct relationship between the optimized electron density $\rho({\bf r})$ 
from Eq.\ (\ref{Shadow_RhoMin}) and the dynamical variable density $n({\bf r})$ we 
may use the equivalent notation $\rho[n]({\bf r}) \equiv \rho({\bf r})$, as above.
The leading difference between the modified potential energy surface ${\cal U}({\bf R},n)$ and 
the regular Kohn-Sham potential $U({\bf R})$, which follows from the linearization,
is of second order with respect to the error in the density.
As long as the dynamical variable density $n({\bf r})$
evolves close to the exact ground state density $\rho_{\rm sc}({\bf r})$, the difference between the modified potential energy 
surface ${\cal U}({\bf R},n)$ and the exact Kohn-Sham potential $U({\bf R})$ will be small. 
This motivates our specific choice of the generalized kernel in Eq.\ (\ref{Kernel}) as will be shown 
in the next section.

\subsection{Ground state optimization kernel}\label{SecKernel}

Our particular choice of the generalized kernel $K({\bf r},{\bf r'})$ 
in Eq.\ (\ref{Kernel}) forces the dynamical variable density $n({\bf r})$
to evolve close to the exact self-consistent ground state density $\rho_{\rm sc}({\bf r})$,
such that the modified Born-Oppenheimer potential energy surface ${\cal U}({\bf R},n)$ is an accurate approximation to
the regular self-consistent Kohn-Sham potential $U({\bf R})$.

Let the functional $F(n)$ be the difference between $\rho[n]({\bf r})$ and $n({\bf r})$, i.e.
\begin{equation}\label{Fn}
{\displaystyle F(n) = \rho[n] - n.}
\end{equation}
The equation $F(n) = 0$ is fulfilled for the exact self-consistent ground state density, 
i.e.\ when $n({\bf r}) = \rho_{\rm sc}({\bf r})$.
If $n({\bf r})$ is sufficiently close to the exact ground state density $\rho_{\rm sc}({\bf r})$, we may
approximate the ground state density through a Newton minimization step \cite{JNocedal99}, 
which schematically is given by
\begin{equation}
{\displaystyle \rho_{\rm sc} \approx n - {\cal J}^{-1} F(n),}
\end{equation}
where ${\cal J}$ is the Jacobian, $\delta F/ \delta n$. This means that
\begin{equation}
{\displaystyle \rho_{\rm sc} - n \approx - {\cal J}^{-1}\left( \rho[n] - n\right) }
\end{equation}
or in a more explicit form
\begin{equation}\label{Newton}
{\displaystyle \rho_{\rm sc}({\bf r}) - n({\bf r}) \approx - \int K({\bf r},{\bf r'})\left( \rho[n]({\bf r'}) - n({\bf r'})\right)d{\bf r'},}
\end{equation}
where the non-local operator $K({\bf r},{\bf r'})$ is the inverse Jacobian of $F(n) = \rho[n] - n$ defined by
\begin{equation}\label{Kernelx}
{\displaystyle \int K({\bf r,r'}) \left(\frac{\delta \rho({\bf r'})}{\delta n({\bf r''})}
- \frac{\delta n({\bf r'})}{\delta n({\bf r''})}\right) d{\bf r'} = \delta({\bf r-r''}) },
\end{equation}
which corresponds to our definition of the kernel in Eq.\ (\ref{Kernel}).
The error in the approximation in Eq.\ (\ref{Newton}), at least under reasonable conditions, 
is of second order, i.e. ${\cal O}\left((\rho - n)^2\right)$,
which can be understood from the quadratic convergence of the Newton scheme for sufficiently
close initial values of $n$ and assuming $F(n)$ is smooth \cite{JNocedal99}.
Our particular choice of the non-local kernel in the generalized extended Lagrangian, Eq.\ (\ref{XLBO}), is thus derived from
an optimization of the extended harmonic potential to accurately mimic the oscillation of $n({\bf r})$
around the exact ground state density $\rho_{\rm sc}({\bf r})$. This improves the accuracy
of the modified potential energy surface ${\cal U}({\bf R},n)$ and helps stabilize the dynamics.

\subsection{Equations of motion}

Previous derivations of the equations of motion in the limit of vanishing self-consistent field
convergence \cite{PSouvatzis13,PSouvatzis14} started from the decoupled equations of motion of extended Lagrangian
Born-Oppenheimer molecular dynamics in the limit of full self-consistent field
convergence.  Then, a simplified set of equations of motion were derived by relaxing the condition 
of the self-consistent field optimization. Such an approach can be difficult 
to interpret. Here we will follow a more direct path in the derivation, without assuming an intermediate
step of full self-consistent field convergence. Instead, we will apply an adiabatic condition that
assumes a separation in the frequencies between the nuclear and the electronic degree of freedom.

The Euler-Lagrange equations of motion for the generalized extended Lagrangian, ${\cal L}({\bf R},{\bf \dot R}, n, {\dot n})$
in Eq.\ (\ref{XLBO}), 
describe a coupled dynamical system of nuclear degrees of freedom with some unknown highest characteristic frequency $\Omega$
and an electronic degrees of freedom with the frequency $\omega$. For normal integration time steps
used in regular Born-Oppenheimer molecular dynamics
we will show in Sec.\ \ref{Sec2F} that $\omega > \Omega$ and that we 
have a natural {\em system-independent} adiabatic separation between the two frequencies.
This natural adiabatic separation between slower nuclear and faster electronic degrees of freedom
motivates our derivation of the equation of motion in the limit when $\omega^2/\Omega^2 \rightarrow \infty$.
In this limit the difference between $\rho({\bf r})$ and $n({\bf r})$
is small and scales as the inverse of the square of the electron frequency $\omega^2$, i.e.\ $(\rho[n] - n) = {\cal O}(\Omega^2/\omega^2)$.
This assumed scaling of $(\rho[n] - n)$ as a function of $\omega^{2}/\Omega^2\rightarrow \infty$ 
is difficult to prove rigorously, but it is straightforward to demonstrate 
that it holds {\em a posteriori} (See Fig.\ \ref{Fig3}).
In this adiabatic limit we further choose to reduce the electron mass parameter 
$\mu \rightarrow 0$ such that $\lim \mu\omega \rightarrow {\rm constant}$.
The Euler-Lagrange equations of motion in this limit are
\begin{equation}\label{G-XLBOMD}\begin{array}{l}
{\displaystyle M_I {\ddot R}_I = - \left. \frac{\partial {\cal U}\left({\bf R},n\right)}{\partial R_I}\right\lvert_n ,}\\
~~\\
{\displaystyle {\ddot n}({\bf r}) = -\omega^2 \int K({\bf r},{\bf r'}) \left(\rho({\bf r'}) - n({\bf r'})\right) d{\bf r'}},
\end{array}
\end{equation}
where the forces in the first equation are given from the partial derivative of ${\cal U}({\bf R},n)$ under the condition
of a constant density $n({\bf r})$, since it is a dynamical variable, and where $K({\bf r},{\bf r'})$ is defined by Eq.\ (\ref{Kernel}).
A more detailed derivation of the equations of motion is given in the appendix.  

The modified Born-Oppenheimer potential energy surface ${\cal U}({\bf R},n)$ is calculated through a single diagonalization, 
which yields the optimized density $\rho({\bf r})$, Eq.\ (\ref{OneStep}). No additional diagonalization of
the Kohn-Sham Hamiltonian is necessary to integrate the equations of motion above 
-- only one Hamiltonian diagonalization per molecular dynamics time step is necessary.  
However, the additional calculation of the kernel $K({\bf r},{\bf r'})$
may introduce a significant overhead.

The key results of this paper are the equations of motion, Eq.\ (\ref{G-XLBOMD}), 
and their derivation from the generalized extended Lagrangian, Eq.\ (\ref{XLBO}),
under the condition of an adiabatic separation between the frequencies
of the nuclear and the electronic degrees of freedom. This allows
stable molecular dynamics simulations for a broader range of materials systems 
than previously possible. The extended range of
systems that can be treated with the new generalized formalism can be understood
from the stability conditions of our previous formulation of extended Lagrangian Born-Oppenheimer
molecular dynamics, as will be discussed in the next section. 

\subsection{Stability}\label{Stability}

It is easy to see that previous formulations of extended Lagrangian Born-Oppenheimer 
molecular dynamics in the limit of vanishing self-consistent field optimization
\cite{ANiklasson12,PSouvatzis13,PSouvatzis14} are recovered
with a scaled $\delta$-function approximation of the kernel, 
i.e.\ for $K({\bf r},{\bf r'}) = -c~\delta({\bf r-r'})$ in Eq.\ (\ref{G-XLBOMD}),
where $c \in [0,1]$. This corresponds to approximating
the exact ground state density $\rho_{\rm sc}({\bf r})$ by a simple linear mixing, i.e. where
\begin{equation}
{\displaystyle \rho_{\rm sc}({\bf r}) \approx c \rho({\bf r}) + (1-c) n({\bf r}),}
\end{equation}
so that 
\begin{equation}
{\displaystyle \rho_{\rm sc}({\bf r}) - n({\bf r}) \approx c \left( \rho({\bf r}) - n({\bf r})\right).}
\end{equation}
The equation of motion for the electronic degrees of freedom with the scaled $\delta$-function approximation of the kernel
in Eq.\ (\ref{G-XLBOMD}),
\begin{equation}
{\displaystyle {\ddot n}({\bf r}) = c \omega^2 \left(\rho({\bf r}) - n({\bf r})\right), }
\end{equation}
therefore corresponds to a dynamics where the auxiliary density $n({\bf r})$ evolves around a
self-consistent ground state density that is approximated by a linear mixing between $\rho[n]$ and $n$.
This linear mixing provides a reasonable approximation of 
the self-consistent ground state density if the electronic energy functional $E({\bf R},{\widetilde \rho})$ 
has a simple convex form, such that
\begin{equation}
{\displaystyle E\left({\bf R},c \rho + (1-c) n\right) < c E\left({\bf R},\rho\right) + (1-c) E \left({\bf R},n\right), ~~~ c \in [0,1].}
\end{equation}
This allows a stable integration of the electronic degrees of freedom in Eq.\ (\ref{G-XLBOMD}) using the
local scaled $\delta$-function approximation of the kernel and requires only one Kohn-Sham diagonalization
per time step.  However, instabilities may occur, 
for example, during bond breaking or in large metallic systems exhibiting charge sloshing.
In this case $\rho({\bf r})$ has to be further updated through an iterative self-consistent field optimization.
Thus, while previous formulations, in the limit requiring only one diagonalization per time step,
rely on stability and convergence using a constant linear mixing, which is possibly guaranteed
only for convex functionals \cite{PHDederichs83}, the generalized formulation with a non-local kernel 
will have a broader range of applications. We can expect that the difference in applicability should be related 
to the range of systems for which a simple constant linear mixing between densities is sufficient 
to reach the self-consistent ground state solution in a regular Kohn-Sham calculation \cite{PHDederichs83}, 
in comparison to problems where Newton-like iterations, such 
a Broyden or Kerker mixing are necessary \cite{CGBroyden65,GPSrivastava84,DDJohnson88,GPKerker81}.

\subsection{Constant of motion}

The constrained optimization of ${\widetilde \rho}({\bf r})$
in Eq.\ (\ref{Shadow_RhoMin}) that defines ${\cal U}({\bf R},n)$ in Eq.\ (\ref{ShadowPot}) and thus the nuclear forces 
in Eq.\ (\ref{G-XLBOMD}), corresponds to a ground state relaxation of the electronic degrees
of freedom with respect to the regular Kohn-Sham energy functional as in Eq.\ (\ref{DFTPot}), 
but with a linearized expression for the electron-electron interaction, Eq.\ (\ref{LinEE}).
In this way ${\cal U}({\bf R},n)$ as defined by Eq.\ (\ref{ShadowPot}) can be understood in terms of
an optimized and thus ${\widetilde \rho}$-independent Born-Oppenheimer potential energy surface with 
a relaxed electron density, but which, thanks to the linearized electron-electron interaction term,
can be obtained through a single diagonalization, Eq.\ (\ref{OneStep}).
We may therefore interpret our one-diagonalization-per-time-step extended Lagrangian Born-Oppenheimer 
molecular dynamics as an exact Kohn-Sham Born-Oppenheimer molecular dynamics simulation,
but with a constant of motion of a modified shadow Hamiltonian, 
\begin{equation}\label{Etot}
{\displaystyle E_{\rm Tot}  = \frac{1}{2}\sum_I M_I{\dot R}_I+ {\cal U}({\bf R},n)},
\end{equation}
which is valid in the same adiabatic limit as the equations on motion, Eq.\ (\ref{G-XLBOMD}).
Thus, instead of integrating the exact Kohn-Sham Born-Oppenheimer equations of motion using an approximate self-consistent
field optimization that leads to a systematic drift in the energy \cite{DRemler90,PPulay04}, we perform the integration based on an exact, fully optimized
ground state density but for an approximate shadow Hamiltonian. 
In this way the energy drift of regular Born-Oppenheimer molecular dynamics can be avoided.
The argument is conceptually very similar to the backward analysis for symplectic integration schemes \cite{RDEngle05}.
As long as $\rho({\bf r})$ and $n({\bf r})$ are close to the self-consistent ground state density $\rho_{\rm sc}({\bf r})$, 
$E_{\rm Tot}$ and the potential energy surface ${\cal U}({\bf R},n)$ 
will be very close to exact Kohn-Sham Born-Oppenheimer molecular dynamics.
This condition is automatically fulfilled when there is an adiabatic separation between $\rho({\bf r})$ and $n({\bf r})$.
As will be shown in the next section, this separation is naturally fulfilled and material independent.

\subsection{Adiabatic separation}\label{Sec2F}

Under the condition of an adiabatic separation between a slower nuclear motion
characterized by a frequency $\Omega$, and a faster auxiliary electronic 
motion governed by the frequency of the extended harmonic oscillator $\omega$, 
we may expect, based on the classical adiabatic theorem, that 
the dynamical variable density $n({\bf r})$, which
is controlled by the harmonic oscillator frequency $\omega$, closely should follow the exact ground state
density $\rho_{\rm sc}({\bf r})$, which is governed by a slower nuclear frequency $\Omega$.
As we will show here, this condition of adiabatic separation between $\Omega$ and $\omega$ is automatically fulfilled.
We can make a straightforward estimate of this separation assuming a standard Verlet integration
of $n({\bf r})$ in the equations of motion in Eq.\ (\ref{G-XLBOMD}),
\begin{equation}\label{Verlet}
{\displaystyle n({\bf r},t+\delta t) = 2n({\bf r},t) - n({\bf r},t-\delta t) - \delta t^2 \omega^2  \int K({\bf r},{\bf r'}) 
\left(\rho({\bf r'}) - n({\bf r'})\right) d{\bf r'} .  }
\end{equation}
It can be shown that the highest possible value of the dimensionless variable
$\kappa = \delta t^2 \omega^2$ that provides a stable Verlet integration is for $\kappa = 2$ \cite{ANiklasson08,ANiklasson09,PSteneteg10}.
For higher order symplectic integration schemes, significantly larger values of $\kappa$ are possible \cite{AOdell09},
though these schemes require multiple force evaluations per time step.
We further assume that the shortest period of the nuclear degrees of freedom, $T_n$ (with the corresponding frequency $\Omega$), 
is integrated in about 20 integration time steps such that $\delta t = T_n/20$,
which is a reasonable choice both in classical and regular Born-Oppenheimer molecular dynamics simulations.
In this case we have that
\begin{equation}
\frac{\Omega}{\omega} = \frac{2\pi/T_n}{\omega} = \frac{2\pi/(20\delta t)}{\sqrt{2}/\delta t} =
\frac{2\pi}{20\sqrt{2}},
\end{equation}
i.e.\ $\omega \approx 4.5 \times \Omega$. Thus, for a normal choice of integration time step
as used in classical or regular Born-Oppenheimer molecular dynamics, we have a 
natural and {\em system independent} separation between the faster
extended harmonic oscillator frequency $\omega$ and the slower nuclear frequency $\Omega$.
It is this natural adiabatic separation that motivates the derivation of the equations
of motion in the adiabatic limit when $\omega^2/\Omega^2 \rightarrow \infty$. The ratio
that governs the error, i.e.\ $\Omega^2/\omega^2 \approx 1/20$, may seem large to motivate
the full adiabatic limit. However, as we will demonstrate in the examples below, this natural 
and material independent separation is sufficient to provide highly accurate molecular dynamics 
simulations.

\section{Examples}\label{Examples}

We have performed an implementation of our generalized extended Lagrangian Born-Oppenheimer dynamics
in the self-consistent charge density functional based tight-binding code LATTE \cite{MCawkwell12,ANiklasson12}. 
It has a particularly simple form that allows a direct calculation of the exact non-local kernel.
In density functional based tight-binding theory \cite{MElstner98,MFinnis98,TFrauenheim00,BAradi07} 
the continuous charge density $n({\bf r})$ is represented
by the atomic net Mulliken charges ${\bf n} = \{n_i\}$ for each atom $i$. 
In the same way, the optimized ground state density $\rho({\bf r})$
is represented by net Mulliken charges ${\bf q} = \{q_i\}$. 
In this formalism the generalized electronic equations of motion in Eq.\ (\ref{G-XLBOMD})
are given by
\begin{equation}
{\displaystyle {\ddot n}_i = -\omega^2 \sum_j K_{ij} \left( q_j - n_j \right) },
\end{equation}
where the kernel is calculated from
\begin{equation}
{\displaystyle \sum_j K_{ij} {\cal J}_{jk} = \delta_{ik}},
\end{equation}
with 
\begin{equation}
{\displaystyle {\cal J}_{jk} = \frac{\delta q_j}{\delta n_k} - \delta_{jk}.}
\end{equation}
The derivatives $\delta q_j/\delta n_k$ have been calculated using 
density matrix perturbation theory \cite{ANiklasson04} and the nuclear degrees of freedom have been integrated
using the velocity Verlet algorithm. For the electronic degrees of freedom we used a modified
Verlet integration scheme \cite{ANiklasson09,PSteneteg10,GZheng11}, where
\begin{equation}\label{ModVerlet}
{\displaystyle n_i(t+\delta t) = 2n_i(t) - n_i(t-\delta t) + \delta t^2{\ddot n}_i(t) + \alpha \sum_{k=0}^K c_k n_i(t-k\delta t).}
\end{equation}
This integration scheme is neither symplectic nor exactly time-reversible when $\alpha > 0$.
However, without the last dissipative force term in the integration scheme, accumulation of numerical noise
may lead to instabilities.

The equations of motion in previous 
formulations \cite{ANiklasson12,PSouvatzis13,PSouvatzis14}
are recovered by a scaled $\delta$-function approximation of the $\delta q/\delta n$ derivative, where
$\delta q_j /\delta n_k \sim \delta_{jk}$. 
In this case the equations of motion are simplified to
\begin{equation}\label{DeltaFuncAppr}
{\displaystyle {\ddot n}_i = c \omega^2  \left( q_i - n_i \right) }, ~~~ c \in [0,1].
\end{equation}
As was previously discussed in Sec.\ \ref{Stability}, this corresponds to a harmonic
oscillator evolving around a ground state density that is approximated by a linear mixing.
As long as linear mixing gives an improved ground state estimate over $n({\bf r})$,
this is a simple, efficient and stable approach. 
For many systems there is therefore no visible effect 
if we switch from the scaled $\delta$-function approximation
to the exact calculation of the non-local kernel in Eq.\ (\ref{Kernel}). However, for material
systems where a simple constant linear mixing is not sufficient to reach a ground state solution
the previous approach will fail. 
For such systems (or if the scaling constant $c$ has been set too large or too small) 
instabilities occur in the time integration of $n({\bf r})$ and the simulations may diverge.
In this case multiple self-consistent field iterations using some more advanced optimization 
scheme are necessary to improve the approximation of
$\rho({\bf r)}$, which significantly increases the computational cost.

\begin{figure}[t]
\resizebox*{5.0in}{!}{\includegraphics[angle=-90]{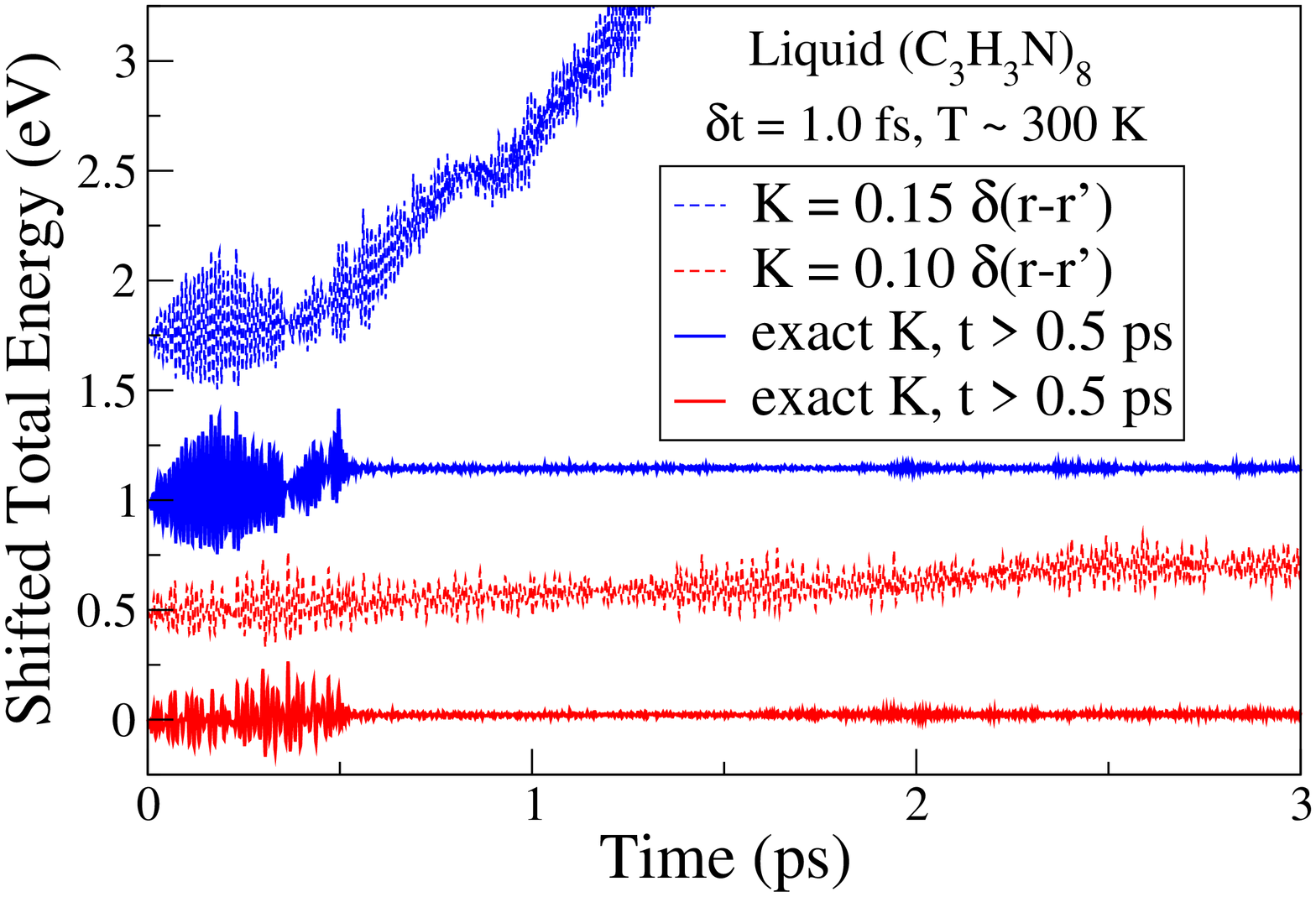}}
\caption{\label{Fig1}
(color online) The fluctuations of the shifted total (kinetic+potential) 
Born-Oppenheimer energy, $E_{\rm Tot}$ in Eq.\ (\ref{Etot}), for 
simulations of liquid acrylinitrile (C$_3$H$_3$N) using
only one diagonalization per time step with various approximations of the kernel, 
$K({\bf r},{\bf r'})$ in Eq.\ (\ref{Kernel}), for
the generalized extended Lagrangian Born-Oppenheimer molecular dynamics, Eq.\ (\ref{G-XLBOMD}).
The two dashed curves illustrate two scaled $\delta$-function approximations of the kernel. 
The two solid lines show the same simulations until $t = 0.5$ ps when we switch from the scaled 
$\delta$-function approximations to the exact non-local kernel.  The exact non-local kernel, Eq.\ (\ref{Kernel}),  
which is calculated at $t = 0.5$ ps and then kept constant for $t > 0.5$ ps, provides 
a very accurate dynamics, which is very close to an exact regular Born-Oppenheimer molecular dynamics.}
\end{figure}

\begin{figure}[t]
\resizebox*{5.0in}{!}{\includegraphics[angle=-90]{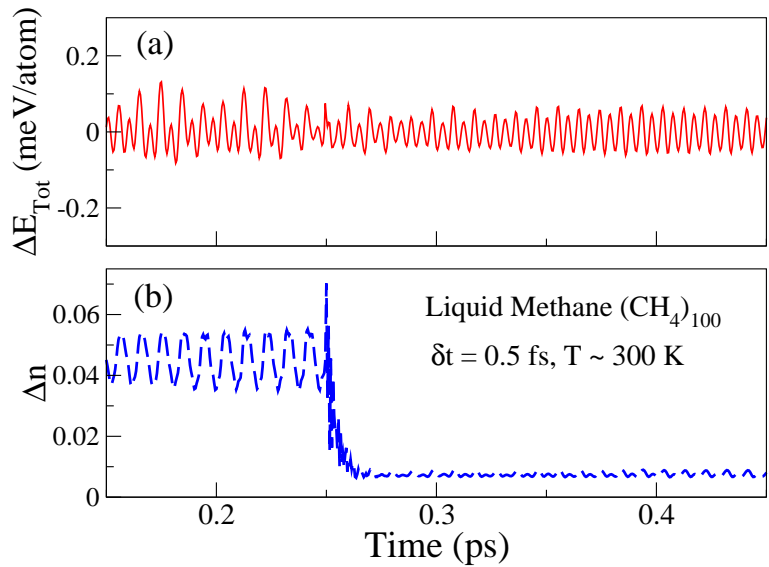}}
\caption{\label{Fig2}
(color online) The upper panel (a) shows fluctuations of the shifted total (kinetic+potential)
Born-Oppenheimer energy, Eq.\ (\ref{Etot}), for a simulation of liquid methane at room temperature. 
The lower panel (b) shows the fluctuations of the 2-norm error
in the charge, $\Delta n = \|n({\bf r}) - \rho_{\rm sc}({\bf r})\|_2$, between 
the auxiliary dynamical variable density and the exactly optimized self-consistent
ground state. The kernel, $K({\bf r},{\bf r'})$,
is approximated with a scaled $\delta$-function, $K({\bf r},{\bf r'}) = -0.25 ~ \delta({\bf r-r'})$, for
$t < 0.25$ ps and with the constant non-local kernel, Eq.\ (\ref{Kernel}),
for $t \ge 0.25$ ps. All calculations of the charges where performed in a single step
directly from the effective Kohn-Sham Hamiltonian, Eq.\ (\ref{OneStep}), using a linear scaling 
density matrix expansion scheme \cite{ANiklasson02,MCawkwell12}.}
\end{figure}

Figure \ref{Fig1} depicts some results of using various approximations to the kernel 
$K({\bf r},{\bf r'})$ in Eq.\ (\ref{Kernel}) for simulations of liquid acrylonitrile.
Only one Hamiltonian diagonalization (or density matrix construction) per time step was used.
Liquid acrylonitrile (C$_3$H$_3$N) was chosen because it shows significant instabilities 
with a dynamics that is hard to stabilize with a simple scaled $\delta$-function 
approximation of the kernel $K({\bf r,r'})$. The two solid curves show what happens
when we switch from a simple scaled $\delta$-function approximation to the exact 
non-local expression in Eq.\ (\ref{Kernel}), which thereafter is kept constant after 
its calculation at $t = 0.5$ ps. The two dashed lines show the
fluctuations in the total energy if we continue using the scaled $\delta$-function approximation.
As is clearly seen, the exact non-local kernel provides a very efficient
stabilization of the dynamics even if it is kept fixed as a constant preconditioner after its initial calculation.
In fact, recalculating the kernel for each new time step (not shown) makes very little difference
to the simulation in Fig.\ \ref{Fig1}. This indicates that it is possible to use more efficient, 
approximate calculations of the kernel in the first place, possibly Broyden 
based schemes \cite{CGBroyden65,GPSrivastava84,DDJohnson88}, 
of Kerker mixing techniques \cite{GPKerker81}, as well as other more recent 
methods \cite{PAnglade08,LLin13} with similar results.
These optimization schemes that are used in a broad range of self-consistent electronic 
structure methods are based on various non-local approximations of the 
exact kernel and reduces the computational cost significantly \cite{JNocedal99}. 

The upper panel of Fig.\ \ref{Fig2} shows the fluctuations in the total energy for a simulation of liquid methane.
The effect on the total energy fluctuations from the exact non-local kernel, which is included at $t = 0.25$ ps, 
is fairly small. However, the error in the charge distribution (lower panel) shows a significant shift, but the error is 
small already at $t < 0.25$ ps when the local scaled $\delta$-function approximation is used for the kernel.
The $\delta$-function approximation was scaled with an artificially small prefactor, $c = 0.25$, to enhance
the effect of the switch to the non-local exact kernel. If a prefactor in the range of $c \in [0.5,0.7]$
had been chosen, the difference would have been much smaller.

The computational cost of 
calculating the exact kernel in Eq.\ (\ref{Kernel}) increases with the number of atoms such that a system
with 100 atoms introduces an additional effort in a single calculation of $K({\bf r,r'})$
corresponding to about 100 normal molecular dynamics time steps without the calculation of $K({\bf r,r'})$. 
Fortunately, the calculation is trivially parallelizable and may thus add very little overhead 
in the total wall-clock time of a simulation, in particular, if the kernel only needs to be 
updated every few thousand time steps. However, for simulations of very large systems, which may require a more
frequent update of the kernel, the full exact calculation of $K({\bf r,r'})$ has to be avoided.

\begin{figure}[t]
\resizebox*{5.0in}{!}{\includegraphics[angle=-90]{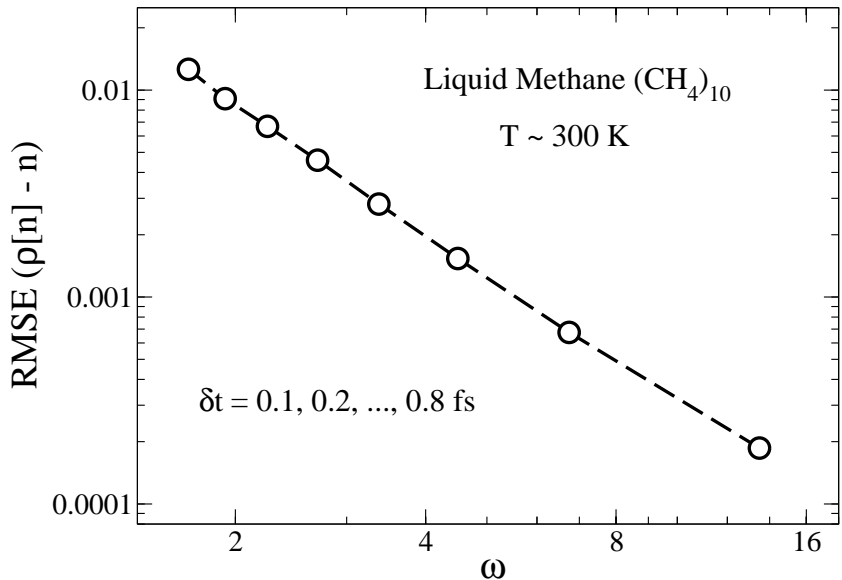}}
\caption{\label{Fig3}
(color online) The root mean square error (RMSE) or the difference between $\rho[n]({\bf r})$ and $n({\bf r})$
in a simulation of liquid methane as a functions of $\omega$. The RMSE scales as $\omega^{-2}$.
The simulations were performed for a sequence of integration time step $\delta t = 0.1, 0.2, \ldots , 0.8$ fs.}
\end{figure}

In Fig.\ \ref{Fig3} we demonstrate the scaling of the difference between $\rho[n]({\bf r})$ and $n({\bf r})$
as a function of $\omega$. The exact non-local kernel was calculated once in the beginning of the simulations 
and was thereafter kept constant.  The root mean
square error between $\rho[n]({\bf r})$ and $n({\bf r})$ was calculated for 2,000 time steps. 
Since $\delta t^2 \omega^2 = \kappa$, which occurs as a constant dimensionless factor 
in the modified Verlet integration, Eq.\ (\ref{Verlet}) or (\ref{ModVerlet}),
we have that $\omega = {\sqrt \kappa }/\delta t$. 
Different values of $\omega$ are thus given by modifying the integration time step $\delta t$.
The figure clearly demonstrates how the difference between $\rho[n]({\bf r})$ and $n({\bf r})$ 
scales as $\omega^{-2}$, which is consistent with the necessary assumption in the derivation of the equations of
motion in the adiabatic limit as $\omega \rightarrow \infty$. This also illustrates the tunable accuracy
of the simulation. Since $\rho[n]({\bf r}) - n({\bf r}) = F(n)$ in Eq.\ (\ref{Fn}), the dynamics converges quadratically
toward exact self-consistent Kohn-Sham Born-Oppenheimer molecular dynamics as $\omega \rightarrow \infty$.
However, already for a normal integration time step when $\delta t  = 0.5$ fs, the root mean 
square error is less than $0.005$.

\begin{figure}[t]
\resizebox*{5.0in}{!}{\includegraphics[angle=-90]{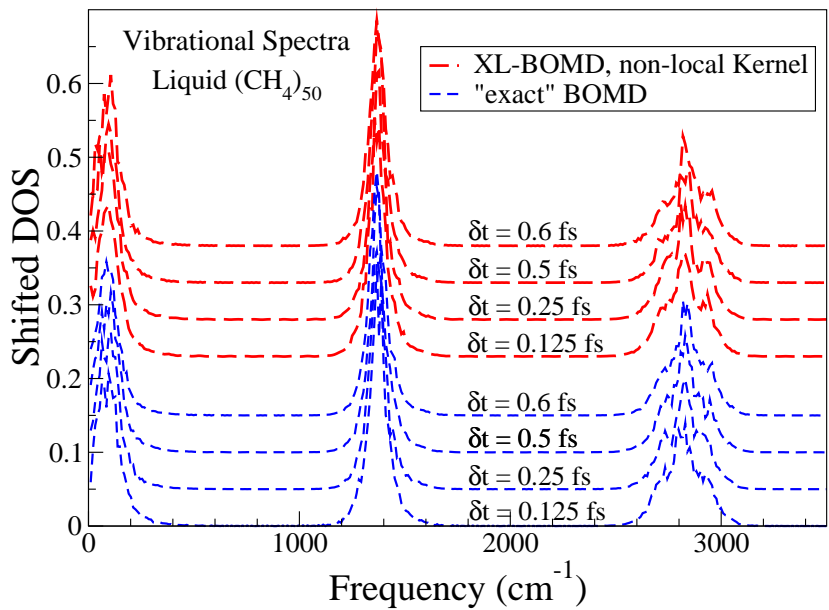}}
\caption{\label{Fig4}
(color online) The vibrational density of states (DOS) for liquid methane
calculated from the velocity autocorrelation function that was sampled 
using either Born-Oppenheimer molecular dynamics with an optimized
ground state density prior to each force evaluation (``exact'' BOMD) 
or the fast generalized extended Lagrangian formulation that requires 
only diagonalization per MD time step (XL-BOMD)
with the optimized non-local kernel $K({\bf r,r'})$ in Eq.\ (\ref{Kernel}).
Four different integration time steps $\delta t$ were used with a temperature
initialized to 300 K during equilibration.}
\end{figure}

The sensitivity to the integration time step and the value of $\omega$
can also be investigated for the vibrational spectra.
Figure \ref{Fig4} shows the (shifted) vibrational density of states (DOS) of liquid Methane given from the
Fourier transform of the velocity autocorrelation function \cite{JMDickey69}. After an initial equilibration 
of 50,000 time steps the velocity autocorrelation was sampled over 100,000 time steps every femtosecond for simulations 
using four different integration time steps $\delta t$. The lower curves show the results
for Born-Oppenheimer molecular dynamics simulations using multiple self-consistent 
field iterations in the optimization of the ground state density before each force evaluation (``exact'' BOMD).
The four upper curves show the same results using our fast generalized extended Lagrangian
formulation that requires only one diagonalization per MD time step
with the optimized non-local kernel. The curves are virtually all on top of each other
without any sensitivity to $\omega$ in for range of integration time steps.
The longest integration time step, $\delta t = 0.6$ ps, is about 1/20 of the shortest
period corresponding to the highest frequency of the system. For time steps $\delta t = 0.75$ ps
we started to se an instability with a small but systematic drift in the energy both for
the optimized Born-Oppenheimer simulation and the extended Lagrangian scheme with 
the non-local kernel.

\section{Discussion}

The generalized extended Lagrangian formulation of Born-Oppenheimer molecular dynamics
allows a more accurate and stable dynamics for a broader class of systems.
Calculations involving bond breaking or charge sloshing, which may occur in simulations 
of chemical reactions or large metallic systems, 
are often not possible to converge using a simple constant linear mixing of densities.
The generalized non-local kernel acting on the electronic degrees of freedom can
avoid such shortcomings and thus significantly increase the application range
of extended Lagrangian Born-Oppenheimer molecular dynamics simulations
in the fast limit that only requires one diagonalization per time step. 

The derivation of the generalized equations of motion based on the adiabatic condition 
has several connections to the analysis of the equations of motion of 
extended Born-Oppenheimer molecular dynamics in its original formulation
that recently was performed by Lin {\em et al.} \cite{LLin14} as well as the
more generalized framework described by Hutter \cite{JHutter12}. Their analysis 
also includes several interesting comparisons and relations to Car-Parrinello molecular 
dynamics \cite{RCar85,MTuckerman96,DMarx00}, which is based on an extended 
Lagrangian approach to first principles molecular dynamics simulations that 
was introduced almost 30 years ago.

The concept of a modified shadow Hamiltonian plays an important role in classical 
molecular dynamics \cite{JPChannel90,McLachlan92,BJLeimkuhler04,RDEngle05}. 
Instead of an approximate numerical integration of an underlying exact
Hamiltonian dynamics, a class of symplectic integration schemes corresponds
to an exact numerical integration of an underlying approximate shadow Hamiltonian. In this
way properties of the dynamical flow of a Hamiltonian dynamics 
can be rigorously conserved. Here we used the concept of a shadow Hamiltonian
in the context of an incomplete self-consistent field optimization
to describe and understand the accuracy in the long-term conservation 
of the total energy.

Although conceptually different and presented in a form that is easy to generalize to other functionals, 
the modified Born-Oppenheimer potential energy surface ${\cal U}({\bf R}, n)$ 
with the linearized electron-electron interaction in Eq.\ (\ref{ShadowPot}) has
the same form as the Harris-Foulkes functional \cite{JHarris85,MFoulkes89}, 
with a second-order error term 
\begin{equation}
{\displaystyle {\cal U}({\bf R},n) - U({\bf R}) = {\cal O}\left((\rho-\rho_{\rm sc})(n-\rho_{\rm sc})\right).}
\end{equation}
The main difference is that ${\cal U}({\bf R}, n)$ is represented in terms of a variationally optimized energy functional
for a given external and electrostatic potential that are determined by the nuclear positions ${\bf R}$ 
and a {\em dynamical variable} density $n({\bf r})$. Since $n({\bf r})$ occurs as a
dynamical variable in the extended Lagrangian, forces that are calculated from the Euler-Lagrange equations
of motion are given by partial derivatives of ${\cal U}({\bf R}, n)$
with respect to a constant density $n({\bf r})$ in Eq.\ (\ref{G-XLBOMD}). This is not possible in a
Harris-Foulkes functional, where $n({\bf r})$ represents either overlapping ${\bf R}$-dependent 
atomic charge densities or the successive (non-converged and thus ${\bf R}$-dependent) 
input densities in a regular iterative Kohn-Sham optimization procedure.
Thus, our modified Born-Oppenheimer potential energy surface ${\cal U}({\bf R},n)$ 
plays a quite different role and allows computationally simple and accurate calculations 
of the forces without relying on the Hellmann-Feynman theorem \cite{DMarx00}.

The important role of $n({\bf r})$ as a dynamical variable is sometimes misunderstood. It is
only thanks to the underlying dynamical framework that our molecular dynamics simulations work. 
Extended Lagrangian Born-Oppenheimer molecular dynamics is sometimes referred to simply 
as a time-reversible extrapolation scheme for the electronic degrees of freedom 
from one step to the next. That picture is misleading in the same way as if we would
view extended Lagrangian Car-Parrinello molecular dynamics simply
as an {\em ad hoc} numerical interpolation technique for wavefunctions.

The calculations of the exact non-local kernel $K({\bf r,r'})$ increases the computational cost
of extended Lagrangian Born-Oppenheimer molecular dynamics. However, for many systems
a local scaled $\delta$-function approximation is sufficient. Other approximate expressions
based on, for example, Broydens method or Kerker mixing could also be applied. In fact, it is
straightforward to invent numerous flavors of local or semi-local approximations
of the Kernel that require different levels of computational complexity.
However, each update of an approximate Kernel, for example based on the Broyden scheme,
would introduce a computational overhead similar to a full self-consistent field optimization.
An understanding of the efficiency of various existing non-local kernel approximations can 
be estimated from their previous applications to accelerate the self-consistent field optimization 
in regular Kohn-Sham density functional theory with a fixed external potential 
\cite{CGBroyden65,PHDederichs83,GPSrivastava84,DDJohnson88,PAnglade08,LLin13}. 
However, there is exist no perfect, fully automatic, black box solution to the self-consistent
field optimization problem and the same limitation can therefore be expected also
for application in the generalized form of extended Lagrangian Born-Oppenheimer 
molecular dynamics.

\section{Summary and conclusions}

We have proposed a generalized framework for extended Lagrangian Born-Oppenheimer
molecular dynamics, Eqs.\ (\ref{XLBO}) and (\ref{G-XLBOMD}), 
that improves the stability and extends the range of applications in the limit of fast quantum based 
molecular dynamics simulations that require only one diagonalization per time step. 
We showed how the equations of motion, Eq.\ (\ref{G-XLBOMD}), 
can be derived directly from the Lagrangian, Eq.\ (\ref{XLBO}), under the condition of an
adiabatic separation between the nuclear and electronic degrees of freedom.
We also showed how this adiabatic separation is automatically fulfilled for
normal choices of the integration time step.

We believe that the general ideas presented in this paper can be directly applied to a broad class
of Born-Oppenheimer-like molecular dynamics simulation techniques based on,
for example, equilibrated or constrained charge relaxation for reactive
or polarizable force field calculations, orbital-free density functional theory,
as well as density matrix, Green's or correlated wavefunction based methods beyond 
the effective Kohn-Sham single particle formulation of density functional theory.

\section{Acknowledgements}

We acknowledge support by the United States Department of Energy (U.S. DOE) Office
of Basic Energy Sciences and the LANL Laboratory Directed Research and Development
Program as well as discussions with B\'alint Aradi, David Bowler, Ondrej Certnik, Eric Chisolm, Petros Souvatzis, 
C.J. Tymczak, Kirill Velizhanin and stimulating contributions by T. Peery at the T-Division Ten Bar Java group.  
LANL is operated by Los Alamos National Security, LLC,
for the NNSA of the U.S. DOE under Contract No. DE-AC52- 06NA25396.

\section{Appendix}

Previous derivations of the equations of motion in the limit of vanishing self-consistent field
convergence followed a different path \cite{PSouvatzis13,PSouvatzis14} from what will be given here. 
In earlier derivations a set of Euler-Lagrange equations of motion for extended Lagrangian 
Born-Oppenheimer molecular dynamics were first derived in the limit of full self-consistent field 
convergence as $\mu \rightarrow 0$.  Then, a new set of equations of motion were derived by relaxing 
the condition of the self-consistent field optimization.  That approach is not fully transparent.
Below we will make a derivation of the equations of motion directly from the generalized extended Lagrangian without
the intermediate step of assuming a full self-consistent field convergence. Instead, we will assume 
an adiabatic separation between the electronic and nuclear degrees of freedom and derive
the equations of motion in the limit as $\omega/\Omega \rightarrow \infty$ and $\mu \rightarrow 0$.
This approach is motivated by the natural adiabatic separation
between the nuclear and the electronic degrees of freedom that was analyzed in Sec.\ \ref{Sec2F}.

The dynamics described by the generalized extended Lagrangian ${\cal L}({\bf R}, {\bf \dot R},n,{\dot n})$
in Eq.\ (\ref{XLBO}) are given by the Euler-Lagrange equations of motion,
\begin{equation}\begin{array}{l}
{\displaystyle M_R {\ddot R}_I = -\frac{\partial {\cal U}}{\partial R_I}
- \frac{1}{2}\mu \omega^2 \frac{\partial}{\partial R_I} \iiint\left(\rho({\bf r}_1)-n({\bf r}_1)\right)^\dagger
K^\dagger({\bf r}_1,{\bf r}_2)K({\bf r}_2,{\bf r}_3)
\left(\rho({\bf r}_3)-n({\bf r}_3)\right)d{\bf r}_1d{\bf r}_2d{\bf r}_3, }\\
~~ \\
{\displaystyle \mu {\ddot n}({\bf r}) = -\frac{1}{2}\mu \omega^2  \frac{\delta}{\delta n}
\iiint\left(\rho({\bf r}_1)-n({\bf r}_1)\right)^\dagger
K^\dagger({\bf r}_1,{\bf r}_2)K({\bf r}_2,{\bf r}_3)
\left(\rho({\bf r}_3)-n({\bf r}_3)\right)d{\bf r}_1d{\bf r}_2d{\bf r}_3  - \frac{\delta {\cal U}}{\delta n}  }.
\end{array}
\end{equation}
For an adiabatic separation between the nuclear frequency $\Omega$ and the electronic frequency $\omega$, 
the difference between $\rho({\bf r})$ and $n({\bf r})$ is small and scales as $ \Omega^2/\omega^2$.
This behavior of the scaling is here asserted without any formal proof. 
However, the situation is very similar to an externally driven harmonic oscillator where the driving
frequency $\Omega$ is small compared to the harmonic oscillator frequency $\omega$. In this case the
$\Omega^2/\omega^2$ scaling can be formally derived \cite{CJ}.
The $\omega^{-2}$ (or the corresponding $\delta t^2$) scaling of $\left(\rho({\bf r}) - n({\bf r})\right)$ 
can consistently be demonstrated {\em a posteriori}, as was shown in Fig.\  \ref{Fig3}. Apart from the scaling
of the difference between $\rho({\bf r})$ and $n({\bf r})$, we can further show that 
\begin{equation}
{\displaystyle \frac{\delta {\cal U}({\bf R},n)}{\delta n} \sim {\cal O}({\rho - n}) \sim \omega^{-2}}.
\end{equation}
This follows from the fact that any variational dependence of $\rho({\bf r})$ with respect to $n({\bf r})$
vanish because of the minimization with respect to ${\widetilde \rho}({\bf r})$ in Eq.\ (\ref{ShadowPot}). 
The only remaining term is the variation with respect to the linearized electron-electron interaction term, 
which scales as ${\cal O}(\rho - n)$, i.e.
\begin{equation}
{\displaystyle\frac{\delta {\cal U}({\bf R},n)}{\delta n} 
\sim \frac{\delta E^{(1)}[n]}{\delta n} 
\sim {\cal O}\left(\rho-n\right) \sim w^{-2}}.
\end{equation}
In the limit when $\mu \rightarrow 0$ and $\omega \rightarrow \infty$ such that $\mu\omega \rightarrow {\rm constant}$
we therefore find that
\begin{equation}\begin{array}{l}
{\displaystyle M_R {\ddot R}_I = -\frac{\partial {\cal U}({\bf R},n)}{\partial R_I}} \\
~~ \\
{\displaystyle {\ddot n}({\bf r}) = - \frac{1}{2} \omega^2  \frac{\delta}{\delta n}
\iiint\left(\rho({\bf r}_1)-n({\bf r}_1)\right)^\dagger
K^\dagger({\bf r}_1,{\bf r}_2)K({\bf r}_2,{\bf r}_3)
\left(\rho({\bf r}_3)-n({\bf r}_3)\right)d{\bf r}_1d{\bf r}_2d{\bf r}_3 }.
\end{array}
\end{equation}
The functional derivative in last equation is more complicated,
\begin{equation}\begin{array}{l}
{\displaystyle {\ddot n}({\bf r}) = - \frac{1}{2} \omega^2  \frac{\delta}{\delta n({\bf r})}
\iiint\left(\rho({\bf r}_1)-n({\bf r}_1)\right)^\dagger
K^\dagger({\bf r}_1,{\bf r}_2)K({\bf r}_2,{\bf r}_3)
\left(\rho({\bf r}_3)-n({\bf r}_3)\right)d{\bf r}_1d{\bf r}_2d{\bf r}_3  } \\
~~\\
{\displaystyle = -\frac{1}{2} \omega^2  \iiint \left( \frac{\delta \rho({\bf r_1}) }{\delta n({\bf r}) } 
- \frac{\delta n({\bf r_1})}{\delta n({\bf r}) } \right)^\dagger
K^\dagger({\bf r}_1,{\bf r}_2)K({\bf r}_2,{\bf r}_3)
\left(\rho({\bf r}_3)-n({\bf r}_3)\right)d{\bf r}_1d{\bf r}_2d{\bf r}_3 } \\
~~\\
{\displaystyle - \frac{1}{2} \omega^2 \iiint \left(\rho({\bf r}_1)-n({\bf r}_1)\right)^\dagger
K^\dagger({\bf r}_1,{\bf r}_2)K({\bf r}_2,{\bf r}_3) 
\left( \frac{\delta \rho({\bf r}_3) }{\delta n({\bf r}) } - \frac{\delta n({\bf r}_3)}{\delta n({\bf r}) } \right)   } \\
~~\\
{\displaystyle - \frac{1}{2} \omega^2 \iiint \left(\rho({\bf r}_1)-n({\bf r}_1)\right)^\dagger
\left( \frac{\delta K^\dagger({\bf r}_1,{\bf r}_2)}{\delta n({\bf r})}K({\bf r}_2,{\bf r}_3) 
+ K^\dagger({\bf r}_2,{\bf r}_1)\frac{\delta K({\bf r}_2,{\bf r}_3)}{\delta n({\bf r})} \right)
\left(\rho({\bf r}_3)-n({\bf r}_3)\right)d{\bf r}_1d{\bf r}_2d{\bf r}_3   }.
\end{array}
\end{equation}
If we assume that $K({\bf r},{\bf r}')$ and $\delta K({\bf r},{\bf r}')/\delta n({\bf r''})$ are bounded, the
last term will vanish in the limit $\omega \rightarrow \infty$, i.e.\ 
\begin{equation}\begin{array}{l}
{\displaystyle {\ddot n}({\bf r}) = -  \omega^2  \iiint \left( \frac{\delta \rho({\bf r_1}) }{\delta n({\bf r}) }
- \delta({\bf r}-{\bf r}_1)  \right)^\dagger K^\dagger({\bf r}_1,{\bf r}_2)K({\bf r}_2,{\bf r}_3)
\left(\rho({\bf r}_3)-n({\bf r}_3)\right)d{\bf r}_1d{\bf r}_2d{\bf r}_3 }. \\
\end{array}
\end{equation}
Using the definition of the kernel corresponding to the transpose of Eq.\ (\ref{Kernel}), 
\begin{equation}\begin{array}{l}
{\displaystyle \int \left( \frac{\delta \rho({\bf r_1}) }{\delta n({\bf r}) }
- \delta({\bf r}-{\bf r}_1)  \right)^\dagger
K^\dagger ({\bf r}_1,{\bf r}_2) d{\bf r}_1 = \delta({\bf r}-{\bf r}_2)},
\end{array}
\end{equation}
we arrive at the final equations of motion for a generalized extended Lagrangian Born-Oppenheimer molecular
dynamics, which is valid under the condition of an adiabatic separation between the frequencies of the
nuclear and the extended electronic degrees of freedom:
\begin{equation}\label{XBO_EQ}\begin{array}{l}
{\displaystyle M_R {\ddot R}_I = -\frac{\partial {\cal U}({\bf R},n)}{\partial R_I},} \\
~~ \\
{\displaystyle {\ddot n}({\bf r}) = 
- \omega^2  \int K({\bf r},{\bf r'}) \left(\rho({\bf r'})-n({\bf r'})\right)d{\bf r'} }. \\
\end{array}
\end{equation}
From these equations of motion it is also easy to understand that the constant of motion, 
which formally would follow from Noether's theorem in the adiabatic limits chosen above, 
is given by Eq.\ (\ref{Etot}). 

%

\end{document}